\documentclass[11pt]{article}
\usepackage{epsfig,Blois}

\def\be{\begin{equation}}
\def\ee{\end{equation}}
\def\bea{\begin{eqnarray}}
\def\eea{\end{eqnarray}}

\begin{document}
\title{HARD POMERON CONTRIBUTION \\ TO FORWARD ELASTIC SCATTERING \footnote{Contribution to
the proceedings of the XIth International Conference on 
Elastic and Diffractive Scattering, Ch\^{a}teau de Blois, 
France, May 15 - 20, 2005.}}

\author{J.R. CUDELL$^{(1)}$, A. LENGYEL$^{(2)}$, E. MARTYNOV$^{(3)}$ and O.V. SELYUGIN$^{(4)}$ }

\address{\small (1) Inst. de Physique, B\^at B5a, Univ. de Li\`ege, B4000 Li\`ege,
Belgium, JR.Cudell@ulg.ac.be;\\
(2)Inst. of Electron Physics, Universitetska 21, UA-88000
Uzhgorod, Ukraine; \\
(3)Bogolyubov Inst.
for Theoretical Physics, UA-03143 Kiev, Ukraine;\\
(4)Bogoliubov Theoretical
Laboratory, JINR, 141980 Dubna, Moscow Region, Russia.\\}
\maketitle\abstracts{The introduction of a hard singularity in fits to
total cross sections and to the ratio of real to imaginary parts
enables to reproduce the data at $\sqrt{s}\leq 100$ GeV
using only simple-pole parametrisations, both for the soft and for the 
hard pomerons.}

\section{Soft pomerons}
Some time ago \cite{COMPETE}, it has been suggested that a pomeron
model à la Donnachie-Landshoff \cite{DL}, 
where the leading singularity in the complex
$j$ plane is given by a simple pole, could not
provide the best fits to the data for total cross sections, and
for the ratio $\rho$ of the real to the imaginary part of the forward 
elastic scattering amplitude. This conclusion was based on an analysis
of all available data at $t=0$ for $\bar p p$, $pp$, $\pi^\pm p$, $K^\pm p$,
$\gamma p$ and $\gamma\gamma$ scattering, where three kinds
of parametrisations were used:\\
$\bullet$ a triple-pole singularity, which makes total cross sections rise
as $\log^2(s)$ at high energy;\\
$\bullet$ a double-pole singularity, which gives $\sigma_{tot}\sim \log(s)$;\\
$\bullet$ a simple-pole singularity, which gives a power rise $s^{\epsilon}$.\\
In each case, non-leading exchanges were accounted for by two simple-pole
contributions contributing to non-degenerate crossing-odd and crossing-even
trajectories.

The conclusions were that
the best fit was always given by the triple-pole
pomeron, closely followed by the double pole, and that the simple-pole
parametrisation was excluded if one went down in energy to $\sqrt{s}=5$ GeV,
or if one included the $\rho$ parameter in the fit. 

In \cite{clms}, we re-examined this question. First of all,
we found that a few sub-leading effects improved the fits significantly:\\
$\bullet$ the use of the theoretical variable, $(s-u)/2\propto\cos\theta_t$ 
rather than $s$, and of the flux factor $2 m_{target}\ \  p^{lab}_{beam}$ instead 
of $s$;\\
$\bullet$ the use of subtraction constants in the dispersion relations giving 
$\rho$ from $\Im m A$.\\
These effects indicate that $\sqrt{s}=5$ GeV is not really in the asymptotic
region: the fit is affected by sub-leading terms, but these are under control.
However, all the fits are improved, so that the triple-pole and the 
double-pole still give a better description of the data, as shown in Table~1. 

However, these two
parametrisations work better for rather peculiar reasons: the dipole becomes
negative below $\sqrt{s}=9.5$ GeV, whereas the tripole has a minimum at 
$\sqrt{s}=5.8$ GeV, and so rises if $s$ decreases. Both of these features
significantly modify
the fit at lower energy, by adding a more complicated $s$
dependance to a power in the $C=+1$ sector. Hence it is natural to check
whether a more complicated $C=+1$ exchange could lead to a better fit
in the simple-pole case. 

Given the hadronic amplitude 
${\cal A}^{ab}$, we define the total cross section as
\begin{equation}
\sigma _{tot}^{ab}\equiv \Im m{\cal A}^{ab}/(2 m_b p_{lab}),
\end{equation}
with $p_{lab}$ the momentum of particle $b$ in the $a$ rest frame,
and the model that we consider is then defined by the following equation:
\begin{equation}
\label{fluxfac}
{\Im}m{\cal A}^{ab}\equiv s_1\left[ {\Im}_{R+}^{ab}\left({\tilde s\over s_1}
\right) +{\Im}_{S}^{ab}\left( {\tilde s\over s_1}\right) \mp {\Im}_-^{ab}
\left({\tilde s\over s_1}\right) \right] ,
\end{equation}
with $s_1=1$ GeV$^2$, and the $-$ sign in the last term for particles,
$ {\Im}_{R+}^{ab}$ and  ${\Im}_{-}^{ab}$ being the contributions of the crossing-even and crossing-odd reggeons.
For the pomeron
contribution ${\Im}_{S}^{ab}$, we allow two simple poles to contribute:
\begin{equation}
\label{poles}
\Im_S^{pb}=S^{b}\left( {\tilde s\over s_1}\right)^{\alpha _{o}}
+H^{b}\left({\tilde s\over s_1}\right)^{\alpha _{H}}
\end{equation}
\begin{table}
\begin{center}
{{\begin{tabular}{|c|c|c|c|c|c|c|}
\hline
Process                       & \( N_{p} \) & 1 simple pole & dipole & tripole&2 simple poles&unitarised \\ \hline
\( \sigma (pp) \)             & 104         & 1.1 & 0.88 & 0.87               &  0.87       &    0.87   \\ \hline
\( \sigma (\bar{p}p) \)       & 59          & 0.88 & 0.94 & 0.94              &  0.92       &    0.92   \\ \hline
\( \sigma (\pi ^{+}p) \)      & 50          & 1.2 & 0.68 & 0.68               &  0.70       &    0.69   \\ \hline
\( \sigma (\pi ^{-}p) \)      & 95          & 0.92 & 0.97 & 0.97              &  0.93       &    0.95   \\ \hline
\( \sigma (K^{+}p) \)         & 40          & 0.97 & 0.73 & 0.71              &  0.72       &    0.72   \\ \hline
\( \sigma (K^{-}p) \)         & 63          & 0.73 & 0.62 & 0.61              &  0.61       &    0.61   \\ \hline
\( \sigma (\gamma p) \)       & 41          & 0.56 & 0.58 & 0.54              &  0.54       &    0.56   \\ \hline
\( \sigma (\gamma \gamma ) \) & 36          & 0.88 & 0.80 & 0.73              &  0.70       &    0.82   \\ \hline
\( \rho (pp) \)               & 64          & 1.6 & 1.6 & 1.7                 &  1.7        &    1.7    \\ \hline
\( \rho (\bar{p}p) \)         & 11          & 0.40 & 0.39 & 0.42              &  0.41       &    0.40   \\ \hline
\( \rho (\pi ^{+}p) \)        & 8           & 2.9 & 1.8 & 1.8                 &  1.6        &    1.7    \\ \hline
\( \rho (\pi ^{-}p) \)        & 30          & 1.9 & 1.0 & 1.0                 &  1.0        &    1.0    \\ \hline
\( \rho (K^{+}p) \)           & 10          & 0.70 & 0.57 & 0.60              &  0.62       &    0.60   \\ \hline
\( \rho (K^{-}p) \)           & 8           & 1.7 & 1.2 & 1.0                 &  0.98       &    1.0    \\ \hline \hline
all, \( \chi ^{2}_{tot} \)    & 619         & 661 & 564 & 558                 &  551        &    557    \\ \hline
all, \( \chi ^{2}/ \)d.o.f.   & 619         & 1.10 & 0.94 & 0.93              &  0.924      &    0.933  \\ \hline
\end{tabular}}}
\end{center}

\caption{Values of the \protect\( \chi ^{2}\protect \) per point for
different processes in fits based on integral dispersion relations
with subtraction constants. The third column corresponds to a simple-pole fit,
the fourth to a double pole, the fifth to a triple pole. The last two
columns show that the inclusion of a hard pomeron leads to a significant
improvement of the fit.}

\end{table}

We give in Table 1 the quality of the new fit. We can see that the inclusion
of this new $C=+1$ singularity
has a dramatic effect: the $\chi^2$ drops from 661 to 551 for
619 points, nominally a 10 $\sigma$ effect! More surprisingly, the
new singularity has an intercept of 1.39, very close to that
obtained in DIS \cite{DLdis}. 

However, as was already known
\cite{C99}, the new trajectory, which we shall call the hard pomeron,
almost decouples from $pp$ and $\bar p p$ scattering. Nevertheless, it improves
considerably the description of $\pi p$ and $K p$ amplitudes, and
parametrisation (\ref{poles}) becomes as good as the tripole fit advocated in
\cite{COMPETE}. This solution is similar to that of ref. \cite{DL04}, where 
only $pp$ and $\bar p p$ scattering were considered. The residue for $pp$ 
is about 20 times smaller than the residues that we get for $\pi p$ and $Kp$,
which seems impossible.

The cause of this suppression is easily understood: 
the hard pomeron generates a 
fast-rising contribution to the cross sections. Hence it must have
a really small coupling to accommodate the data at the highest energies.
In the pion and kaon cases, the data extend only to 63 GeV, so that a
relatively large coupling is allowed. This leads to two conclusions:
first of all, to study the hard contribution, one needs to limit oneself
to low energy. We chose to limit all data to $\sqrt{s}=$ 100 GeV.
Secondly, to describe the S$p\bar p$S and Tevatron data, one needs to
unitarise the hard pomeron contribution.

Following this program, we obtained the parameters for the hard and soft
pomerons which are summarised in Table 2.
Two main conclusions can be drawn: the hard pomeron intercept is
\begin{equation}
\alpha_H(0)=1.45\pm0.01
\end{equation}
and the couplings to protons, pions and kaons are
\begin{equation}
H_p:H_\pi:H_K=1:2.8:3
\end{equation}
The origin of this hierarchy remains mysterious, although it could be a size
effect, the hard pomeron being a short-distance effect.

Note that in this fit, we have used Regge factorisation for the couplings
of the photon:
\begin{equation}
H_{\gamma\gamma}=H_{\gamma p}^2/H_{pp};\ 
S_{\gamma\gamma}=S_{\gamma p}^2/S_{pp}
\end{equation}
and similar equations for the reggeon exchanges. Hence one can conclude\cite{DL04} that the hard pomeron obeys factorisation, although the quality of
the $\gamma\gamma$ data leaves this point unsettled.
\begin{table}
\begin{center}
{\begin{tabular}{|c|c|c|c|c|} \hline & \multicolumn{2}{c|}{ soft+hard poles }& \multicolumn{2}{c|}{soft pole+ unitarised hard}\\ \hline
Parameters & value & error & value & error \\ \hline
\( \alpha _{o} \) & 1.0728 & 0.0008 & 1.0728 & fixed \\ \hline
\( S_{p} \)& 56.2 & 0.3 & 55 & 1 \\ \hline
\( S_{\pi } \) & 32.7 & 0.2 & 31.5 & 0.9 \\ \hline
\( S_{K} \)& 28.3 & 0.2 & 27.4 & 0.8 \\ \hline
\( S_{\gamma } \)& 0.174 & 0.002 & 0.174 & 0.003 \\ \hline
\( \alpha _{h}(0) \) & 1.45 & 0.01 & 1.45 & fixed \\ \hline
\( G_{p} \)& -- & -- & 0.18 & 0.06 \\ \hline
\( G_{\gamma } \) & -- & -- & 6\( \times 10^{-9} \) & 1.5\( \times 10^{-8} \) \\\hline
\( H_{p} \)& 0.10 & 0.02 & 0.17 & 0.05 \\ \hline
\( H_{\pi } \)& 0.28 & 0.03 & 0.43 & 0.08 \\ \hline
\( H_{K} \)& 0.30 & 0.03 & 0.42 & 0.07 \\ \hline
\( H_{\gamma } \)& 0.0006 & 0.0002 & 0.0005 & 0.0002 \\ \hline
\end{tabular}}
\end{center}
\caption{Parameters obtained in the fits. The second and third columns give
the parameters and errors of the fit with a hard pole, Eq. (\ref{poles})
for \protect\( \sqrt{s}\protect \) from 5 to 100 GeV, the fourth
and fifth columns give the parameters of a unitarised fit, Eq.
(\ref{unit})
for 5 GeV\protect\( <\sqrt{s}<2\protect \) TeV. }
\end{table}

Finally, one needs to tackle the issue of unitarisation. 
Indeed, the hard singularity cannot be extended
to energies beyond a few hundred GeV, where one reaches the black-disk
limit.
The problem of course is that nobody knows how to unitarise Regge
exchanges unambiguously. 
We know that if 1-pomeron exchange is given by the amplitude
\[
\Im mA(s,t)\approx g_{1}\left( \frac{s}{s_{1}}\right) ^{\alpha _{H}}\,
e^{R_H^{2}t}\]
with $
R_H^{2}=B_H+\alpha_H '\log s$
then, if the hadrons remain intact during multiple exchanges, the
\( n \)-pomeron contribution at $t=0$ will be proportional to

\[
\Im mA^{(n)}(s)\propto (-1)^{n-1}\, s\, \frac{s^{n(\alpha _{H}-1)}}{\left[
R_H^{2}\right] ^{n-1}} \]
The coefficients of the successive terms, and the influence of 
triple pomeron vertices are unknown.
In order to show that it is possible to reproduce the data via unitarisation,
we chose \cite{clms} the simplest form, which can be obtained in the $U$-matrix formalism, 
which reproduces simple-pole exchange at small $s$, and obeys the
Froissart-Martin bound at high $s$:
\begin{equation}
\label{unit}
\Im mA^{H}_+(s)=H_{a}sR^{2}\left[ \frac{1}{G}\log \left\{ 1+G\frac{s^{\alpha
_{h}-1}}{R^{2}}\right\} \right] .
\end{equation}
with $B_H=4$ GeV$^2$ and $\alpha'$ =0.1 GeV$^{-2}$. 
We also assumed that $G$ would be the same
for all hadrons, and different for photons, and fixed the hard pomeron intercept to that obtained in the previous fit. We show in Table 1 that such a form
can reproduce the Tevatron data well, while being almost identical to a simple-pole at low energy, $G$ being of the order of 20 \%. Table 2 gives
the parameters corresponding to this unitarised form. 

In conclusion, it could well be that the hard singularity that was predicted
30 years ago by BFKL was already present then in the data. It may have
been observed in DIS, and it would be very surprising
that no trace of it would subsist in soft data. Of course, if it is associated
with short-distance fluctuations, then it can appear only rarely. But its inclusion does help the description of the data at $t=0$. The hierarchy of
its couplings is unexpected, and it seems compatible with the factorisation
properties of a simple pole. Its presence in elastic cross sections
has also now been motivated by the next contribution to these proceedings \cite{clm}, and it could well modify significantly the total cross
section at the LHC \cite{LHC}.

\end{document}